\documentclass[prd,aps,floats,nofootinbib,showpacs,preprintnumbers]{revtex4}

\usepackage{amsmath}
\usepackage{amssymb}
\usepackage{graphicx}
\usepackage[dvips]{color}

\def\be{\begin{equation}}
\def\ee{\end{equation}}
\def\bea{\begin{eqnarray}}
\def\eea{\end{eqnarray}}
\def\a{\alpha}

\def\om{\omega}

\def\p{\partial}

\def\lag{{\cal L}}

\newcommand{\pp}{~~~.}
\newcommand{\vv}{~~~,}
\definecolor{Black}{named}{Black}
\definecolor{Red}{named}{Red}

\begin{document}

\setlength{\unitlength}{1mm}

\title{Relaxing Nucleosynthesis Constraints on Brans-Dicke Theories}

\author{Antonio De Felice\footnote{{\tt A.De-Felice@sussex.ac.uk}}$^{1,2}$,
Gianpiero Mangano\footnote{{\tt mangano@na.infn.it}}$^{3}$, Pasquale
D. Serpico\footnote{{\tt serpico@mppmu.mpg.de}}$^{4}$, Mark
Trodden\footnote{{\tt trodden@phy.syr.edu}}$^{2}$}

\affiliation{$^1$Department of Physics \& Astronomy, University of Sussex, Brighton BN1 9QH, UK.\\
$^2$Department of Physics, Syracuse University, Syracuse, NY
13244-1130, USA.\\
$^3$INFN, Sezione di Napoli, and Dipartimento di Scienze Fisiche
Universit\`a di Napoli Federico II Complesso Universitario di Monte
S.\ Angelo, Via Cintia, I-80126 Napoli, Italy\\
$^4$Max-Planck-Institut f\"{u}r Physik (Werner-Heisenberg-Institut),\\
F\"{o}hringer Ring 6, D-80805 Munich, Germany}

\begin{abstract}
We reconsider constraints on Brans-Dicke theories arising from the
requirement of successful Big Bang Nucleosynthesis. Such constraints
typically arise by imposing that the universe be radiation-dominated
at early times, and therefore restricting the contribution that a
Brans-Dicke scalar could make to the energy budget of the universe.
However, in this paper we show how the dynamics of the Brans-Dicke
scalar itself can mimic a radiation-dominated kinematics, thereby
allowing successful nucleosynthesis with a sizable contribution to
the total cosmic energy density. In other words Newton's constant
may dynamically acquire values quite different from that today, even
though the evolution mimics radiation domination. This
possibility significantly relaxes the existing bounds on Brans-Dicke
fields, and opens the door to new possibilities for early universe
cosmology. The necessary fine tunings required by such an
arrangement are identified and discussed.
\\
\begin{center}
{\it Dedicated to Rafael Sorkin, on the occasion of his 60th
birthday, \\ and to celebrate his wonderful contributions to
physics}
\end{center}

\end{abstract}

\maketitle

\section{Introduction}
In the last few years, there have been a large number of different
approaches to the dark energy enigma. As is well-known, the
anisotropies of the Cosmic Microwave Background (CMB)
radiation~\cite{Spergel:2006hy} and data from type Ia
Supernovae~\cite{Sn1,Sn2,Sn3}, are well-fit by a cosmological
constant, albeit a fine-tuned one, of which we have no
satisfactory theoretical understanding. Another possibility is
that the cosmological constant is precisely zero (or at most
subdominant) and that a dynamical component is driving cosmic
acceleration. This component may be a manifestation of a
large-scale modification of General Relativity
\cite{Carroll:2003wy,Carroll:2004de,Capozziello:2003tk,Deffayet:2001pu,Freese:2002sq,
Arkani-Hamed:2002fu,Dvali:2003rk,Arkani-Hamed:2003uy}, or a new
dynamical field, such as a scalar field
\cite{Wetterich:1987fm,Ratra:1987rm,Caldwell:1997ii,Wang:1999fa,Sahlen:2005zw}.
It is quite natural to consider scalar fields that are
non-minimally coupled to gravity
\cite{bd1,Brans:1961sx,Dicke:1961gz,Carroll:2004hc,Perivolaropoulos:2003we,
Faraoni:2001tq,Gunzig:2000kk,Riazuelo:2001mg,Uzan:1999ch,Perrotta:2002bk,Perrotta:2002sw,
Perrotta:1999am,Pettorino,Pettorino:2005pv,Chiba:2001xx,Chen:1999qh,Torres:2002pe,Santos:1996jc}.
Such fields arise both in the context of string theory, and in
more general theories with extra spatial dimensions. These fields
are typically of the Brans-Dicke (BD) type, and have been
considered in a large variety of other cosmological contexts, such
as inflation and baryogenesis
\cite{DeFelice:2004uv,Chen:2005tr,Feng:2004mq,Li:2004hh,Mathiazhagan:1984vi,
Barrow:1990nv,Steinhardt:1990zx,Green:1996hk,Liddle:1991am,Crittenden:1992cf,La:1989za,Berkin:1990ju}.

However, there exist tight constraints on BD fields. Some of these
constraints are on solar system scales, constraining the BD
parameter to satisfy $\om>40000$. Other constraints arise from
cosmology itself. For example, in order not to spoil the success of
the Big Bang Nucleosynthesis (BBN) predictions, the scalar field
component should be subdominant to radiation and the scale factor
must evolve nearly as $t^{1/2}$. This constraint is quite well
accepted, and a detailed analysis of the constraints on other kinds
of evolutions, such as dark matter dominated one, can be found
in~\cite{Carroll:2001bv}.

In this paper we revisit this particular issue, considering
whether a BD scalar field might play a {\it significant} role in
cosmic evolution during BBN, while maintaining an acceptable
evolution of the scale factor. We shall only require the BD field to
drive the kinematics of the universe, making it expand {\it as if}
radiation dominates. It should be emphasized that in our model the
expansion rate (Hubble function $H$) is driven by the BD field,
while the scattering rates are determined by cross sections and
abundances of the standard fields of the plasma ($e^{\pm}$,
neutrinos, photons, baryons). One consequence is that the relation
between temperature and time becomes an adjustable parameter in this
model. Provided the BD field can reproduce a standard kinematics
$H(a)$, one is free e.g.\ to normalize the abundances of all the
fields to the value they assume at the same $H(a)$ in the standard
model. Since they do not interact with the BD field but
gravitationally, the particles in the plasma cannot distinguish
which component is driving the evolution of the universe, and they
will interact among themselves in the same way as if the BD field is
not present. As a consequence, the production rate for the light
elements is the same and the freeze-out temperatures for the
reactions are unchanged.

The big difference between this model and other BD models during BBN
(see e.g. \cite{Coc:2006rt}), is that now the BD field,
i.e~Newton's constant, can be largely different from today's value
at the time of BBN. This is quite different from a standard approach
in which the BD field is allowed to be just slightly different from
today's value. However, just after BBN, some non-trivial dynamics is
required to bring the BD fields to values consistent with a standard
evolution for the universe. This is sufficient to ensure that the
temperature of the universe has the usual behavior from the end of
BBN up to now. We shall see that such a scenario is possible, while
remaining consistent with solar system constraints. However, as one
might expect, this interesting possibility can only arise under
finely tuned conditions. We describe the tuning required, both on
the initial conditions of the scalar and on its associated
potential.

\section{Scalar tensor theories}

We consider the BD Lagrangian density with a potential
$W(\varphi)$ as
\begin{equation}
\lag=\sqrt{-g}\left[\varphi\,R-\frac{\om}{\varphi}\,(\partial_\mu\varphi)^2-W(\varphi)+16\pi\,\lag_{\rm
m}\right] \vv
\end{equation}
where $R$ is the Ricci scalar, $\varphi$ is a real scalar field
with units of [mass]$^2$, $\om$ is the BD parameter and $\lag_{\rm
m}$ denotes the matter Lagrangian density. Our signature is
${-}{+}{+}{+}$.  Typically, as long as the
potential is too weak to confine the scalar field, precision
measurements of the timing of signals from the Cassini mission
yield the bound $\om > 40000$ \cite{cassini}. Although it has been
suggested that this bound may not hold on cosmological scales
\cite{bound}, we will take a conservative view and demand that it
be satisfied.

Varying the action with respect to the metric tensor gives
\begin{equation}
G_{\mu\nu}-\frac{\om}{\varphi^2}\,\p_\mu\varphi\,\p_\nu\varphi-\frac{1}{\varphi}\,\nabla_\mu\nabla_\nu\varphi+
g_{\mu\nu}\left[\frac{\Box\varphi}{\varphi}+\frac{\om}{2\,\varphi^2}\,(\p\varphi)^2+\frac{
W}{2\,\varphi}\right] =\frac{8\pi}{\varphi}\,T_{\mu\nu} \vv
\end{equation}
and combining this result with the equation obtained by varying
the action with respect to $\varphi$ one obtains
\begin{equation}
\Box\varphi=\frac{8\pi}{2\om+3}\,T+\frac{1}{2\om+3}\,[\varphi\,W_{,\varphi}-2\,
W] \vv
\end{equation}
where $T={T_{\mu}}^\mu$ and $W_{,\varphi}$ denotes $d W/d\varphi$.

To study the cosmological aspects of this theory in the framework of a
homogeneous and isotropic universe, we use the Friedmann,
Robertson-Walker (FRW) ansatz for the metric
\begin{equation}
ds^2=a^2(\eta)\left(-d\eta^2 +d{\bf x}^2\right) \vv
\end{equation}
where $a(\eta)$ is the scale factor and $\eta$ is conformal time,
related to the usual cosmic time through $a(\eta)d\eta=dt$. We have
assumed a zero curvature contribution, consistent with current data.

If we introduce the dimensionless field $\phi\equiv
\varphi/M_P^2$--- $M_P$ being the Planck mass---the equations of
motion become
\begin{eqnarray}
\ddot{\phi}&=&-2\,\frac{\dot{a}}{ a}\,\dot{\phi}+ \frac{8\pi\,
a^2}{(2\om+3)M_P^2} \left[\rho-3\,P+\frac{
W}{4\pi}-\frac{\phi}{8\pi}\,
\frac{d W}{d\phi}\right] \vv \label{eqzNS1}\\
\frac{ \dot{a}}{ a}&=&-\frac{\dot{\phi}}{2\phi}+
\left[\frac{2\om+3}{12}\left(\frac{\dot{\phi}}{\phi}\right)^{\!2}+
\frac{8\pi\,a^2}{3M_P^2\,\phi}\left(\rho+\frac{W}{16\pi}\right)\right]^{\!1/2}
\vv \label{eqzNS2}\,
\end{eqnarray}
where $\rho$ and $P$ refer to the total energy density and
pressure of the standard plasma components respectively, and a dot
denotes a derivative with respect to conformal time $\eta$.

\section{Mimicking Radiation During Big Bang Nucleosynthesis}
In the standard scenario, the universe is radiation dominated
during BBN. This requirement leads to theoretical expectations
that are in reasonable agreement with observations, at least at the 2
$\sigma$ level. In particular the Deuterium number
density~\cite{tytler} points to a value for the baryon content of
the universe which is within 1-$\sigma$ equal to the independent
determination of this parameter extracted from the CMB power
spectrum~\cite{Spergel:2006hy}
\begin{equation}
\Omega_b h^2 =0.0223 \pm 0.0008 \pp \label{omcmb}
\end{equation}
This result is compatible with a radiation content given by
photons and three weakly interacting neutrinos/antineutrinos (for
a review of the present status of standard BBN see
e.g.~\cite{cuocoetal,serpicoetal,cyburt}).

Nevertheless, present data still leave room for a non-standard
energy density content, or a non standard value for the Hubble
expansion rate. For example, the analysis carried out in
\cite{serpicoetal} shows that a conservative observational range
for the helium-4 mass fraction constrains any extra
contribution or deficit in the radiation energy content in the
range $-1.1 \lesssim \Delta N_{\rm eff} \lesssim 0.8$, when
parameterizing it in terms of extra effective neutrino species.

In view of these considerations, the main question we would like to
address is whether it is still possible that the BD field $\phi$
contributes during BBN in a non negligible way. The effect of $\phi$
can be easily discerned from Eq.~(\ref{eqzNS2}). In particular, its
dynamics affects the value of the gravitational constant which
scales as $1/ \phi$ and which ties the energy-density to the
expansion rate. Moreover, the dynamics of $\phi$ provides extra
contributions to the Hubble parameter. The usual assumption is to
consider a negligible role of the functions $W/16 \pi \phi$ (the
``potential energy term'') and $K \equiv (2 \omega+3) M_P^2
(\dot{\phi}/\phi)^2/32 \pi a^2$ (the ``kinetic energy term'') in
Eq.~(\ref{eqzNS2}) \footnote{Notice that this definition of the
kinetic and potential energy density for $\phi$ is somehow
imprecise, since their sum does not correspond in general to a
covariantly conserved energy density. Nevertheless they are useful
quantities to quantify the contribution of the BD field dynamics to
the expansion rate and moreover, their sum is covariantly conserved
in the limit of very large $\omega$ and negligible energy density
$\rho$ of standard matter which we discuss in this section, see e.g.
\cite{Perrotta:2002sw}.}, so that the usual BBN scenario is only
possibly changed by a different value of the effective gravitational
constant or, equivalently, of a re-scaled energy density, see
e.g.~\cite{damour}. The same bound previously quoted in term of
$N_{\rm eff}$ then translates into $0.9\alt\phi\alt1.2$.

We now relax the hypothesis that the $\phi$ field dynamics
contributes negligibly at the epoch of BBN. Instead, we consider
the possibility that the behavior of $\phi$ during the temperature
interval $1$ MeV $\geq T\geq$ $0.01$ MeV may be highly nontrivial
and yet may still result in primordial abundances for $^2$H,
$^3$He, $^4$He and $^7$Li of the correct order of magnitude. In
particular, we consider a specific, yet interesting case, in which
the universe during this stage is still {\it radiation} dominated,
but by this we now only mean that the specific dependence of the scale
factor $a(t)\propto t^{1/2}$, or equivalently in terms of the
conformal time
\begin{equation}
a(\eta)=\kappa \, \eta \vv \label{radia}
\end{equation}
where $\kappa$ is a constant to be evaluated by cosmological
observables. The condition (\ref{radia}) in turn imposes severe
constraints on Eqs.~(\ref{eqzNS1}) and~(\ref{eqzNS2}), which under
this ansatz can be recast into
\begin{eqnarray}
\phi''&=&-\frac{2}{\eta}\phi'+\frac{8\pi\,a^2}{\kappa^2\,M_P^2\,(2\om+3)}
\left[\rho-3\,P+\frac{W}{4\pi}-\frac{\phi}{\phi'}\,\frac{W'}{8\pi}\right] \vv \label{first}\\
\frac{1}{a}&=&-\frac{\phi'}{2\,\phi}+\left[\frac{2\om+3}{12}\left(\frac{\phi'}{\phi}\right)^2+
\frac{8\pi\,a^2}{3\,\kappa^2\,M_P^2\,\phi}\left(\rho+\frac{W}{16\pi}\right)\right]^{1/2}
\vv \label{pot}
\end{eqnarray}
where a prime denotes a derivative with respect to $a$. Squaring
Eq.~(\ref{pot}) one obtains
\begin{equation}
W=-16\pi\rho+\frac{\kappa^2\,M_P^2}{\phi\,a^4}
\left[6\,\phi^2+6\,a\,\phi\,\phi'-\om\,a^2\,\phi'^2\right] \pp
\label{pott}
\end{equation}
Substituting this expression for $W$ into Eq. (\ref{first}) yields
the following nonlinear ordinary differential equation for $\phi$
\begin{equation}
\phi''=\frac{4\,\phi^2+2\,a\,\phi\,\phi'-a^2\om\,{\phi'}^2}{a^2\,\phi}
+\frac{8\pi\,a^2}{\kappa^2\,M_P^2\,(2\om+3)}
\left[2\frac{\phi}{\phi'}\,\rho'-3(\rho+P)\right] \pp
\label{norma}
\end{equation}
Note that by writing the differential equation in this form we
have excluded the case in which the field satisfies the first
order equation
\begin{equation}
\frac{1}{a}=-\frac{\phi'}{2\,\phi} \vv
\end{equation}
which from Eq. (\ref{pot}) implies that the associated solution is
only possible for a negative potential $W$. Therefore, we shall
not consider this case further.

Using the covariant conservation of the energy for standard matter
$a \rho'=-3(\rho+P)$ , Eq.~(\ref{norma}) can be further simplified
into
\begin{equation}
\phi''=\frac{4\,\phi^2+2\,a\,\phi\,\phi'-a^2\om\,{\phi'}^2}{a^2\,\phi}
-\frac{24\pi\,a(\rho+P)}{\kappa^2\,M_P^2\,(2\om+3)}\left(a+2\frac{\phi}{\phi'}\right)
\pp \label{norma2}
\end{equation}

We are interested in a situation in which the expansion rate of
the universe is of the order of the standard value leading to a
successful BBN, and yet the $\phi$ contribution is relevant or
even dominant. By inspection of (\ref{eqzNS2}) we see that in
order for these conditions to hold, the energy density $\rho$
should be suppressed by a value of $\phi>1$. Furthermore, the new
terms, which are dependent on the dynamics of $\phi$, should {\it
mimic} a radiation dominated behavior and compensate for the
reduced contribution of ordinary radiation to the expansion rate.
In order to see whether this scenario is in fact possible at all,
we start by considering the simplest case, in which ordinary
radiation and matter contribute negligibly to the
kinematics. Of course, this means that we have to check {\it a
posteriori} that, during BBN, it is possible to choose values of
$\phi$ large enough so that $\rho$ is sufficiently suppressed. The
leading contribution to~(\ref{eqzNS2}) will then be provided by
the dynamics of $\phi$.

\begin{figure}[t]
\includegraphics[width=8truecm]{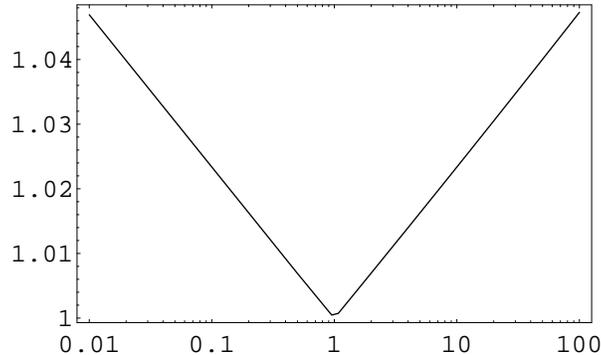}
\caption{The evolution of the $\phi/\Phi_0$ field versus $a$ for
$C=0.5$ and $\omega=40000$. The field bounces back at $a=1$ where
$\phi=\Phi_0$.} \label{fig:bounce}
\end{figure}

If we thus neglect the second term at the r.h.s of
Eq.~(\ref{norma2}), the replacement $s=\phi^{1+\om}$ linearizes the
equation, giving
\begin{equation}
s''=\frac{2}{a}s'+\frac{4\,(\om+1)}{a^2}\,s \vv
\end{equation}
from which the general solution for $\phi(a)$ can be expressed as
\begin{equation}
\phi(a) =\Phi_0\left[(1-C)\,a^{(3-\Delta)/2}+C\,a^{(3+\Delta)/2}
\right]^{1/(1+\om)} \vv \label{campo2}
\end{equation}
with $\Delta=\sqrt{25+16\,\om}$, $\Phi_0$ a normalization
constant giving the value of the field at $a=1$, and $C$
determines the initial value of $\phi'$ via
\begin{equation}
\phi'(1)=\frac{\Phi_0}{2 (1+\omega)} \left[ \Delta(2 C-1)+3
\right] \pp
\end{equation}
It should be noticed that in order that the solution remain valid
in a neighborhood of the BBN epoch we should have $0 \leq C \leq
1$. Furthermore, it can be checked that the potential $W$ fails to
be positive definite if $C>1$. There are two possible monotonic
behaviors for $\phi$ in the two particular cases $C=0$ and $C=1$
\begin{equation}
\phi_\pm(a)=\Phi_0\,a^{\epsilon_{\pm}} \vv \label{phia}
\end{equation}
where $\epsilon_{\pm}\equiv(3\pm\Delta)/2(1+\omega)=\pm 8/(\Delta
\mp 3)$. The $\epsilon_+$ and $\epsilon_-$ indexes correspond to the
field rolling towards larger or smaller values respectively, with an
extremely slow power-law behavior since, for large $\omega$, we have
$|\epsilon_{\pm}|\simeq 2/\sqrt{\om}<0.01$. For all other choices of
$C$ the solution satisfies a very finely tuned condition
corresponding to the field bouncing back at $\Phi_0$, as shown in
Figure~\ref{fig:bounce}. In the following therefore, we will only
consider the two behaviors described in Eq.~(\ref{phia}).
Notice that since $\omega$ is very large, in the interval $0.01 \leq
a \leq 100$ the two solutions can be expressed with an accuracy
better than $1 \%$ as
\be \phi_\pm (a)= \Phi_0 \left(1 \pm
\frac{2}{\sqrt{\omega} }\log a \right) \pp
\ee

Plugging Eq.~(\ref{phia}) into Eq.~(\ref{pott}) (and neglecting
$\rho$) one has
\begin{equation}
\frac{W}{\phi}=\kappa^2\,M_P^2(6+6\epsilon_{\pm}-\om\epsilon_{\pm}^2)
\left(\frac{\phi}{\Phi_0}\right)^{-4/\epsilon_\pm} \simeq 2
\kappa^2\,M_P^2 \left(\frac{\phi}{\Phi_0}\right)^{ \pm 2
\sqrt{\omega}} \vv \label{pot2}
\end{equation}
where the last expression holds for large $\omega$.

Let us summarize the results obtained so far. By considering a
potential of the form shown in Eq.~(\ref{pot2}), the expansion
rate of the Universe due to a BD field behaves the same way as during a
radiation dominated regime. Depending on the sign of the
power-index in $W$, there are two possible behaviors for the field
$\phi$, increasing or decreasing with cosmic expansion. The $\phi$
field rolls very smoothly along the potential, and with comparable
``kinetic'' and ``potential'' energy densities $K$ and $W/\phi$
respectively.  Since we derived these results in the limit $W\gg
\rho$, we must check that this condition is compatible with the
solution found. Indeed, it is easily seen that the term $W/\phi$
in Eq.(\ref{pot2}) is independent from the value of $\Phi_0$,
while the term $\rho/\phi$ can be made arbitrarily small by
choosing $\Phi_0$ arbitrarily large. This implies that provided we
choose the initial value for $\phi$ to be sufficiently large, the
contribution of ordinary radiation can be neglected, and the
expansion of the Universe follows $a \propto \eta$ despite the
fact that it is driven by a BD scalar field.

\section{BBN in a BD-dominated cosmology}
In the previous section we have found that there are two solutions
for which a BD field dominates cosmic dynamics, while keeping a
radiation-dominated expansion as in Eq. (\ref{radia}). We turn
now to the phenomenological requirement of preserving in this
scenario the general agreement between BBN predictions and the
observed light element yields. Of course, the scenario outlined in
the previous section of a pure $\phi$-dominated universe cannot
describe the late stages of cosmic evolution. Indeed this would be
at variance with the fact that today we require $\phi=1$.
Moreover, with the potential considered in the previous section,
the universe would continue to expand as in a radiation dominated
regime until very small redshifts. We do not address these
issues further here, but it is clear that a more complicated dynamics
after the BBN period is required in order to bring such kind of
models in accordance with ``late'' cosmological and astrophysical
observables. In the following, in particular we will assume that
CMB observations are explained as in the standard scenario.

The set of equations governing BBN, besides
equations~(\ref{eqzNS1}) and~(\ref{eqzNS2}), is the following (see
e.g.~\cite{serpicoetal}) \bea
\frac1{n_B}\frac{d n_B}{dt}&=&-3H\\
\frac{d\rho}{dt}&=&-3H(\rho+P)\\
\frac{dX_i}{dt}&=&\Gamma_i(X_j)\\
L(m_e/T,\Phi_e)&=&\frac{n_B}{T^3}\,\sum_j Z_j\,X_j\\
(\partial_t-H\,|\vec p|\,\partial_{|\vec p|}) f_{\nu_\a}(|\vec p|,
t)&=&I_{\nu_\a}[f_{\nu_e},f_{\bar\nu_e},\dots]\ .
\eea
The first
two equations state respectively the conservation of the total
baryon number and energy density in a comoving volume. The third
equation is the Boltzmann equation which describes the density
of each nuclide species, with the $\Gamma_i$ being the rates of
interaction averaged over the kinetic equilibrium distribution
functions. The fourth equation accounts for the electric
charge neutrality of the universe in terms of the electron chemical potential
$\Phi_e$. Finally, the last equation is the Boltzmann equation for
the neutrino species.

It is clear that the field $\phi$ never explicitly enters these
equations but does appear in the evolution of the Hubble parameter.
Therefore, as long as the evolution of $H$ in the BD case is
indistinguishable from the standard case, adopting initial
conditions (at whatever $T\agt 1\,$MeV) for the matter and radiation
fields analogous to the standard case, the production of the light
elements will also be unchanged. However, even in the scenario
described in the previous section there is one relevant departure,
since the Hubble factor evolves strictly as $a^{-2}$, while in the
standard case this evolution is modified because $e^\pm$
annihilation changes the number of effective relativistic degrees of
freedom. Note that this phenomenon happens just between the two
crucial temperatures for BBN; $T\simeq 0.7\,$MeV at which the
neutron/proton ratio freezes, and $T\simeq 0.07\,$MeV when the
deuterium bottleneck opens and light element production begins. If
the BD Hubble rate matches one of the two conditions, it might fail
to match the second one. To test if such a difference can be
accommodated at all, we have run the BBN code described in
\cite{serpicoetal} by assuming that the dominant contribution to $H$
is provided by the BD field, and its behavior is as for pure
radiation as discussed in the previous section. To this end we
replace the standard matter/radiation energy density $\rho$ in the
Hubble law (and only there!) with the following parametrization in
terms of a single neutrino energy density
\begin{equation}
H \rightarrow \sqrt{\frac{8\pi}{3M_P^2}\,y \rho_\nu }\vv
\label{effnu}
\end{equation}
where $\rho_\nu=7  \pi^2 T_\nu^4/120$ and $T_\nu \propto a^{-1}$. In
the limit in which the solution previously found holds, this useful
parametrization is {\it exact}. We ask if there is any value at all
for the factor $y$ which can accommodate at least $^4$He and $^2$H
abundances. Since we assume that the standard cosmology holds for
$a\gg a_{\rm BBN}$, we fix the baryon abundance to that
deduced from the CMB, Eq.~(\ref{omcmb}). Note that, since there are
two conditions to fulfill with only one parameter, the constraint is
non-trivial. We find that, for
\begin{equation}
5.5\lesssim y\lesssim 7.4 \vv \label{boundy}
\end{equation}
the conservative constraint on the $^4$He mass fraction $0.232\leq
Y_p\leq0.258$ quoted in~\cite{Olive:2004kq} is satisfied, while
predicting a deuterium fraction $1.99 \cdot 10^{-5}\leq$ $^2$H/H
$\leq 2.66 \cdot 10^{-5}$, comparable with the observed one
$^2$/H$=(2.78^{+0.44}_{-0.38} ) \cdot 10^{-5}$ \cite{tytler} within
2$\,\sigma$. For $y$ in the range of Eq. (\ref{boundy}) we also
found $0.97 \cdot 10^{-5}\leq$ $^3$He/H $\leq 1.07 \cdot 10^{-10}$,
which is consistent with present bounds \cite{bania}, and
$4.2 \cdot 10^{-10}\leq$ $^7$Li/H $\leq 5.1 \cdot 10^{-10}$, which
remains a factor $\sim$ 3 larger than the observed one
\cite{ryan,b20031}.

In terms of $\phi$, this constraint translates into
\begin{equation}
2.35\sqrt{\frac{8\pi\,\rho_\nu}{3M_P^2}}\leq
-\frac{\dot{\phi}}{2a\,\phi}+
\left[\frac{2\om+3}{12}\left(\frac{\dot{\phi}}{a\,\phi}\right)^{\!2}+
\frac{W}{6\,M_P^2\,\phi}\right]^{\!1/2} \leq
2.72\sqrt{\frac{8\pi\,\rho_\nu}{3M_P^2}} \vv
\end{equation}
or equivalently
\begin{equation}
5.16\, (T_\nu\,a)^2\leq\kappa\,M_P\left(
\frac{\epsilon_{\pm}}{2}+\sqrt{1+\epsilon_{\pm}+\frac{\epsilon_{\pm}^2}{4}}
\right)\leq 5.97\, (T_\nu\,a)^2 \pp \label{resfin}
\end{equation}
The existing bound on $\omega$ implies $|\epsilon_{\pm}|<0.01$, which in
turn simplifies the previous constraint to the following one (with a percent accuracy)
\begin{equation}
5.16\, (T_\nu\,a)^2\leq\kappa\,M_P \leq 5.97\, (T_\nu\,a)^2 \pp
\end{equation}
This is the constraint on the potential $W$---or, more precisely, on
$W/\phi$---of Eq. (\ref{pot2}) which accommodates our model with
successful nucleosynthesis.  From current cosmological measurements
one would deduce $T_{\nu}\,a\simeq 2\,$K and constant from the BBN
epoch till now while. However, lacking a detailed modelling of the
transition from a pure $\phi$-dominated phase to a radiation
dominated one, we do not elaborate on this bound further\footnote{In
the ``standard cosmology'', the value of $\kappa$ would be $$ \kappa
= H\,a^2=H_0 \, a_0^2 \sqrt{\Omega_r},$$ where $a_0$ and $\Omega_r$
are the scale factor and the radiation fraction today including also
neutrinos, independently of the fact that at least some of them are
non-relativistic today. However, unless one specifies the mechanism
for the field $\phi$ to exit the epoch of $\phi$-dominance, one has
no way to relate the $\kappa$ needed at the BBN epoch with
present-day cosmological quantities.}.

\begin{figure}[t]
\includegraphics[width=8truecm]{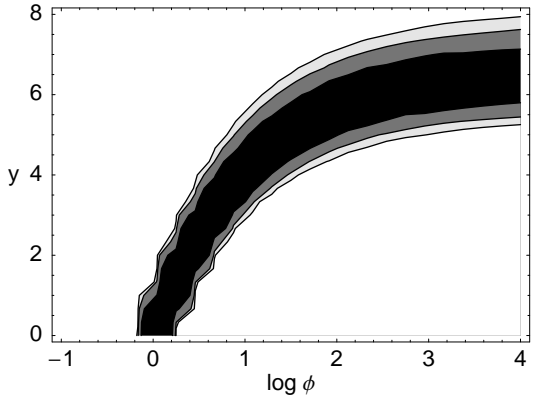}
\caption{The 1 (black), 2 (grey) and 3$\sigma$ (light grey)
contours in the $\phi-y$ plane.} \label{fig:likely}
\end{figure}

 Since we have shown that BBN
is compatible not only with a ``minimal" effect, but also with a
major role for a BD field, it is reasonable to expect that
intermediate cases in which the contribution of $\phi$ and of
ordinary radiation are comparable might be viable. Of course, such
scenarios generally involve a high degree of fine tuning in the
parameters of the field $\phi$ and/or in the choice of the potential
$W$. Thus, for illustrative purposes, in the following we shall take
a phenomenological approach and discuss the situation where
Eq.~(\ref{eqzNS2}) is replaced by
\begin{equation}
\frac{ \dot{a}}{ a^2}=H\simeq
\left[\frac{8\pi}{3M_P^2}\left(\frac{\rho}{\phi}+ \frac{W}{16 \pi
\phi} + K
\right)\right]^{\!1/2}\rightarrow\left[\frac{8\pi}{3M_P^2}\left(\frac{\rho}{\phi}+y\,\rho_\nu\right)\right]^{\!1/2}
\vv \label{bothcase}
\end{equation}
and again consider the case for which the BD dynamics corresponds to a pure
radiation behavior, but relaxing the hypothesis of a vanishing
contribution of ordinary radiation to the Hubble parameter. In the
first equality of Eq.~(\ref{bothcase}) we have neglected the term
$\dot{\phi}/2\,a\,\phi$, which is anyway suppressed by a factor
$1/\sqrt{\omega}<0.01$. Running the modified code for this case
leads to the bounds shown in Figure~\ref{fig:likely}, where we plot
the likelihood contours in the $\phi-y$ plane at 1, 2 and 3-$\sigma$
level using both $^2$H and $^4$He experimental results as before.
Two limits are easily recognized. For $\phi\gg 1$ we recover the
case considered previously of a dominant BD contribution to $H$,
and correspondingly the bound on $y$ reported in Eq. (\ref{boundy}),
while for $y\ll 1$, we obtain the ``trivial" bound on the effective
Newton constant during BBN, namely $0.9\alt\phi\alt 1.2$. This
result shows that not only both regimes are viable, but also that
intermediate situations are phenomenologically allowed, although the
details of these cases are more model dependent, and the parameters
in the BD sector require a high degree of fine tuning.

It should be noticed that to impose the condition that BD behaves as radiation, or that
\begin{equation}
K+\frac W{16\pi\phi}=y\,\rho_\nu\ ,
\label{potenza}
\end{equation}
corresponds to the choice of a particular potential.
In principle one should show that this choice is possible. In order to see if this is indeed sensible, we use~(\ref{potenza})
 in equations (\ref{eqzNS1})--(\ref{eqzNS2}) and find that a solution can be found by taking the time derivative of
these two equations. The outcome is a system of nonlinear 2$^{\rm
nd}$ order differential equations in $a$ and $\phi$. There is only
one independent initial condition---the choice of $\phi_i$---since
$\dot a_i$ can be found by using equation~(\ref{eqzNS2}),
 and since equation~(\ref{eqzNS1}) becomes a cubic equation in $\dot\phi$.
This means that the system becomes uniquely determined and the solution does indeed exist and it is unique.

\section{Comments and conclusions}\label{conclusions}
We have revisited the constraints on a Brans-Dicke scalar field,
arising from the requirement of successful nucleosynthesis. This is
an important question in light of the different roles that a
Brans-Dicke field $\phi$ may play in various sectors of cosmology,
such as inflation, baryogenesis and dark energy.

Indeed, we have found that the BD field, i.e.~Newton's constant, can
have significantly different values from today's one, and at the
same time dominate the energy density making the universe behave as
if radiation dominates. The matter content is not changed at all, so
that all physical quantities such as freeze-out temperatures and
decay/production rates for elements do not differ from standard
ones.

In other words, we have found that it is possible for the BD field
to mimic the effect of radiation on the evolution of the cosmic
scale factor. In such a case $\phi$ is the dominant contributor to
the Hubble parameter during BBN. We have put constraints on the BD
potential using the observed light elements yields and the BBN
predictions in such a modified scenario. This scenario requires a
large initial value for $\phi \agt 10$ so that the standard
contribution of matter to the expansion rate is suppressed by the
reduced value of the effective Newton constant, as well as potential
for the BD field $W$ that scales as a power law in $\phi$, \be
W/\phi \propto \phi^{\pm 2\sqrt\om}\ . \ee Here $\om \geq 40000$ if
we assume that the local result of the Cassini mission also holds on
cosmological scales. We do not offer any suggestion how to explain
such an high exponent. By numerically solving the set of BBN
equations we have found that it is indeed possible to tune the
overall scale of the potential $W$, see Eq. (\ref{resfin}), so as to
obtain final yields for both $^2$H and $^4$He compatible with the
observed data.

Of course, one may worry that large values of $\phi$ at the BBN
epoch are not compatible with today's value. In fact, if we
considered the same effective potential valid until today, the field
$\phi$ would keep on being the dominant component of the universe,
leading to a completely different evolution for the universe, e.g.
an extrapolation would lead to value for the field $\phi$
incompatible with today's value for the Newton gravitational
constant. Therefore, after BBN, the potential should be chosen such
that the field decreases to acceptable values, and the universe
follows a standard evolution. The fact that today's value of $\phi$
imposes an upper bound on the possibility of having very large
values for $\phi_{\rm BBN}$ is tantamount to saying that a
fine-tuning of the choice of the potential is necessary in order to
allow a huge change in the value of the field from BBN up to now.

Finally, we have considered the more general scenario where both
standard radiation and the BD field contribute to the expansion
rate, in the case that the BD dynamics again corresponds to a pure
radiation behavior. In this case too, by comparing the theoretical
prediction with light nuclei abundances, we have found that,
depending on the value of the BD field during BBN and its
contribution to $H$, there is a region in this two-parameter space
leading to successful nucleosynthesis. This region shown in
Figure~\ref{fig:likely} extends from the high $\phi$ regime already
discussed down to the standard scenario with $\phi=1$ showing that
provided the $\phi$ potential is suitably tuned, basically all
values of the BD field can be shown to be compatible with the
abundances of both $^2$H and $^4$He produced during BBN.

An interesting feature of the scenario considered in this
paper is that one is basically free to adjust the time-temperature
relation, which analytically means rescaling the initial conditions
for the thermodynamical quantities of the fluids in the plasma. We
have limited our discussion to the case where $H(a)$ around the BBN
epoch has the same evolution it would have in a radiation-dominated
plasma, keeping the particle densities fixed at the same value of
the standard case as our initial conditions. This is not a
fine-tuning required a priori, but an useful working hypothesis
which authomatically ensures that the rates for the scattering
processes (including the nuclear reactions) are unchanged. However,
one may relax this assumption, and thus the allowed region in the
model space may enlarge considerably, to the price of major (and
model-dependent) modifications in the physical processes in the
plasma. We do not discuss further this point in our phenomenological
approach, since our purpose in this article was to show that exotic
scenarios where the BD field is driving the expansion rate at the
BBN epoch are indeed possible.  Nonetheless, this feature should be
kept in mind when characterizing specific models (as e.g.
quintessence potentials with solutions tracking radiation) with a
cosmic evolution before and after the BBN epoch determined
self-consistently within the theory.

\acknowledgments

G.M.~acknowledges full support by Syracuse University during his
visit to the Physics Department, Syracuse University. MT and ADF are
supported in part by the National Science Foundation, under grant
NSF-PHY0354990, by funds from Syracuse University and by Research
Corporation. ADF is supported in part by PPARC. PS is supported, in
part, by the Deutsche Forschungsgemeinschaft under grant No.~SFB 375
and by the European Union under the ILIAS project, contract
No.~RII3-CT-2004-506222.

\end{document}